\documentclass[twoside]{article}
\usepackage{fleqn,espcrc2}
\usepackage{graphicx}

\newcommand{\AmS}{{\protect\the\textfont2
  A\kern-.1667em\lower.5ex\hbox{M}\kern-.125emS}}

\title{
Two-flavor lattice QCD study of the axial charges of
N(1535) and N(1650)}

\author{Toru T. Takahashi\address{Yukawa Institute for Theoretical Physics, 
        Kyoto University, \\ 
        Kitashirakawa-Oiwakecho, Kyoto 606-8502, Japan}%
        and 
        Teiji Kunihiro\address{Department of Physics, Kyoto University, \\
        Kitashirakawa-Oiwakecho, Kyoto 606-8502, Japan}}
       
\begin{document}

\begin{abstract}
We show the first lattice QCD results on the axial charges $g_A^{N^*N^*}$
of $N^*(1535)$ and $N^*(1650)$.
The measurements are performed with two flavors of
dynamical quarks employing the renormalization-group improved gauge
action at $\beta$=1.95 and the mean-field improved clover quark action
with the hopping parameters, $\kappa$=0.1375, 0.1390 and 0.1400. 
We construct 2$\times$2 correlation matrices and diagonalize them
so that the signals of $N^*(1535)$ and $N^*(1650)$
are properly separated.
Wraparound contributions in the correlator, which can be another source
of signal contaminations, are eliminated by imposing the Dirichlet
boundary condition in the temporal direction. 
The axial charge of $N^*(1535)$ is found to 
take small values as $g_A^{N^*N^*}\sim {\mathcal O}(0.1)$,
whereas that of $N^*(1650)$ is approximately 0.5,
which is almost independent of quark masses
and consistent with the predictions by the naive nonrelativistic quark model.
\end{abstract}

\maketitle

\section{Introduction}
Chiral symmetry together with its spontaneous breaking
is one of the key ingredients 
in the low-energy hadron or nuclear physics.
Due to its spontaneous breaking, up and down quarks,
whose current masses are of the order of a few MeV,
acquire the large constituent masses of a few hundreds MeV,
and are consequently responsible for about 99\% of mass
of the nucleon and hence that of our world.
One would thus consider that chiral condensate $\langle \bar \psi \psi \rangle$,
the order parameter of the chiral phase transition,
plays an essential role in the 
hadron-mass genesis in the light quark sector.
On the other hand, chiral symmetry gets restored 
in systems where hard external 
energy scales such as high-momentum transfer,
temperature($T$), baryon density and so on exist,
owing to the asymptotic freedom of QCD.
Then, several questions may arise:
Are all hadronic modes massless in such systems?
Can hadrons be massive even without non-vanishing chiral condensate?

An interesting possibility was suggested 
some years ago by DeTar and Kunihiro~\cite{DeTar:1988kn},
who showed that nucleons can be {\it massive even without
the help of chiral condensate}
due to the possible {\it chirally invariant mass terms},
which give {\it degenerated}
finite masses to the members in the chiral multiplet
(a nucleon and its parity partner)
even when chiral condensate is set to zero:
In order to demonstrate this possibility
for a finite-$T$ case, they introduced 
a linear sigma model which 
has a nontrivial chiral structure in the baryon sector
and a mass-generation mechanism
essentially different from that 
by the spontaneous chiral symmetry breaking.
Interestingly enough, their chiral doublet model has recently become
a source of debate as a possible scenario of
{\it observed parity doubling in excited baryons
}~\cite{Glozman:2007jt,Jaffe:2005sq,Jaffe:2006jy,Jido:1999hd,Jido:2001nt,Lee:1972},
although their original work \cite{DeTar:1988kn} was 
intended for an application to finite-$T$ systems.

It is thus an intriguing problem to clarify the chiral structure
of excited baryons in the light quark sector
beyond model considerations.
One of the key observables which are sensitive to
the chiral structure of the baryon sector is axial charges~\cite{DeTar:1988kn}.
The axial charge of a nucleon $N$ is encoded in the three-point function
\begin{equation}
\langle N|
A_\mu^a
|N\rangle
=
\bar u
\frac{\tau^a}{2}
[
\gamma_\mu \gamma_5
g_A(q^2)
+
q_\mu \gamma_5
h_A(q^2)
]
u.
\end{equation}
Here, 
$A_\mu^a
\equiv
\bar Q \gamma_\mu \gamma_5 \frac{\tau^a}{2} Q$
is the isovector axial current.
The axial charge $g_A$ is defined by $g_A(q^2)$
with the vanishing transferred momentum $q^2=0$.
It is a well-known fact
that the axial charge $g_A^{NN}$ of $N(940)$ is 1.26.
Though the axial charges in the chiral broken phase
can be freely adjusted with higher-dimensional possible terms
and cannot be the crucial clues for the chiral 
structure~\cite{Jaffe:2005sq,Jaffe:2006jy},
they would reflect the internal structure of baryons
and would play an important role in the clarification of the low-energy
hadron dynamics.

In this paper, 
we show the first unquenched lattice QCD study~\cite{Takahashi:2007ti}
of the axial charges $g_A^{N^*N^*}$ of $N^*(1535)$ and $N^*(1650)$,
two lowest nucleon resonances in the negative parity channel.
We employ $16^3\times 32$ lattice with two flavors of dynamical quarks,
generated by CP-PACS collaboration~\cite{AliKhan:2001tx}
with the renormalization-group improved
gauge action and the mean-field improved clover quark action.
We choose the gauge configurations at $\beta=1.95$
with the clover coefficient $c_{\rm SW}=1.530$,
whose lattice spacing $a$ is determined as 0.1555(17) fm.
We perform measurements with 590, 680, and 680 gauge configurations
with three different hopping parameters for sea and valence quarks,
$\kappa_{\rm sea},\kappa_{\rm val}=0.1375,0.1390$ and $0.1400$,
which correspond to quark masses of $\sim$ 150, 100, 65 MeV and 
the related $\pi$-$\rho$ mass ratios are
$m_{\rm PS}/m_{\rm V}=0.804(1)$, $0.752(1)$ and $0.690(1)$, respectively.
Statistical errors are estimated by the jackknife method
with the bin size of 10 configurations.

\section{Formulation}

Our main concern is the axial charges of the negative-parity 
nucleon resonances $N^*(1535)$ and $N^*(1650)$ in $\frac12^-$channel.
We then have to construct an optimal operator
which dominantly couples to $N^*(1535)$ or $N^*(1650)$.
We employ the following two independent nucleon fields,
\begin{equation}
N_1(x)\equiv \varepsilon_{\rm abc}u^a(x)(u^b(x)C\gamma_5 d^c(x))
\end{equation}
and
\begin{equation}
N_2(x)\equiv \varepsilon_{\rm abc}\gamma_5 u^a(x)(u^b(x)C d^c(x)),
\end{equation}
in order to construct $2\times 2$ correlation matrices
and to separate signals of $N^*(1535)$ and $N^*(1650)$.
Here, $u(x)$ and $d(x)$ are Dirac spinors for u- and d- quark,
respectively, and $a,b,c$ denote the color indices.
Even after the successful signal separations,
there still remain several signal contaminations:
Signal contaminations by {\it scattering states}
and {\it wraparound effects}.

Due to the unquenched gauge configurations,
the negative parity nucleon states could decay to $\pi$ and N,
and their scattering states could inevitably get into the spectrum.
The sum of the pion mass $M_\pi$ and the nucleon mass $M_N$
is however in our setups heavier than the masses of the lowest two states
(would-be $N^*(1535)$ and $N^*(1650)$)
in the negative parity channel.
We then do not suffer from any scattering-state signals.
The other possible contamination is wraparound 
effects~\cite{Takahashi:2005uk}.
Since we perform unquenched calculations,
the excited nucleon $N^*$ can decay into $N$ and $\pi$,
and even when we have no scattering state $|N+\pi\rangle$,
we could have another type of ``scattering states''.
The baryonic correlator
$\langle N^*(t_{\rm snk}) \bar N^*(t_{\rm src})\rangle$
can still accommodate, for example, the following term.
\begin{eqnarray}
&&\langle \pi|N^*(t_{\rm snk})|N\rangle
\langle N| \bar N^*(t_{\rm src})|\pi\rangle \nonumber \\
&\times&e^{-E_N(t_{\rm snk}-t_{\rm src})}\times
e^{-E_\pi (N_t-t_{\rm snk}+t_{\rm src})}.
\end{eqnarray}
Here, $N_t$ denotes the temporal extent of a lattice.
Such a term is quite problematic and mimic a fake plateau
at $E_N-E_\pi$ in the effective mass plot
because it behaves as $\sim e^{-(E_N-E_\pi)(t_{\rm snk}-t_{\rm src})}$.
In order to eliminate such contributions,
we impose the Dirichlet condition on the temporal
boundary for valence quarks,
which prevents valence quarks from going over the boundary.
(Wraparound effects can be found even in quenched 
calculations~\cite{Takahashi:2005uk}.)

We employ zero-momentum-projected point-type operators
for the sinks, and
employ wall-type operators in the Coulomb gauge for the sources.
After we diagonalize the $2\times 2$ correlation matrices,
we can construct optimized operators for N(1535) and N(1650) states.
Once we construct optimized operators,
we can easily compute the (non-renormalized) vector and axial charges
$g_{V,A}^{\pm{\rm [lat]}}$ for the positive- and negative-parity nucleons
via three-point functions with the so-called sequential-source 
method~\cite{Sasaki:2003jh}.
In practice, we evaluate $g_{A}^{\pm{\rm [lat]}}(t)$ defined as
\begin{equation}
g_{A}^{\pm{\rm [lat]}}(t)
=
\frac
{
{\rm Tr}\ \Gamma_{A}
\langle B(t_{\rm snk})
J_\mu^{A}(t)
\overline B(t_{\rm src}) \rangle
}
{
{\rm Tr}\ \Gamma_{A}
\langle B(t_{\rm snk})
\overline B(t_{\rm src}) \rangle
},
\end{equation}
and extract $g_{A}^{\pm{\rm [lat]}}$ by the fit
$g_{A}^{\pm{\rm [lat]}}=g_{A}^{\pm{\rm [lat]}}(t)$ in the plateau region.
$B(t)$ denotes the (optimized) interpolating field for nucleons.
$\Gamma_{A}$ is $\gamma_\mu \gamma_5 \frac{1+\gamma_4}{2}$, and
$J_\mu^{A}(t)$ is an axial vector current inserted at $t$.

\begin{figure}[h]
\begin{center}
\includegraphics[scale=0.38]{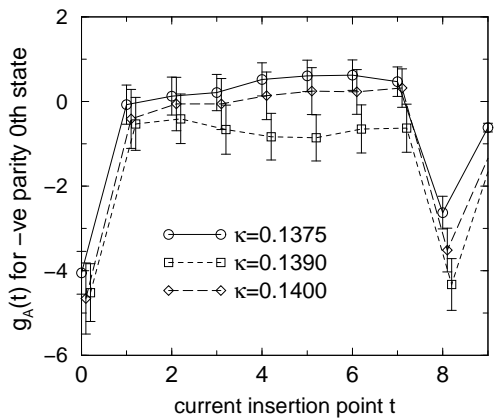}
\end{center}
\vspace{-1cm}
\caption{\label{3pointfunc}
The non-renormalized axial charge of $N^*(1535)$,
$g_{A}^{-0{\rm [lat]}}(t)$,
as a function of the current insertion time $t$.
}
\end{figure}

We show in Fig.~\ref{3pointfunc}
the non-renormalized axial charge $g_{A}^{-0{\rm [lat]}}(t)$ for $N^*(1535)$
as a function of the current insertion time $t$.
They are rather stable around $t_{\rm src}<t<t_{\rm snk}$.

We finally reach the renormalized charges
$g_{A}^\pm=\widetilde Z_{A}g^{\pm{\rm [lat]}}_{A}$
with the prefactors
$\widetilde Z_{A}
\equiv
2\kappa
u_0
Z_{A}
\left(
1+b_{A}\frac{m}{u_0}
\right)
$,
which are 
estimated with the values listed in Ref.~\cite{AliKhan:2001tx}.

\begin{figure}[h]
\begin{center}
\includegraphics[scale=0.38]{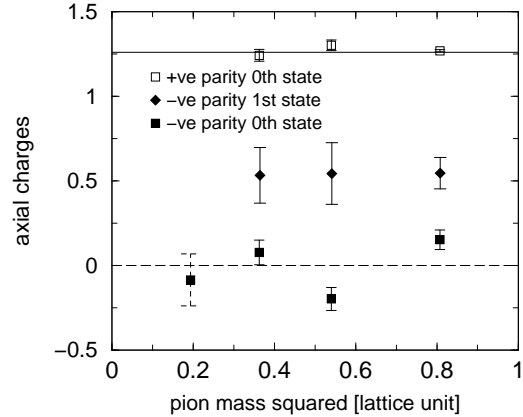}
\end{center}
\vspace{-1cm}
\caption{\label{AxialVectorC}
The renormalized axial charges of the positive- and the 
negative-parity nucleons are plotted
as a function of the squared pion mass $m_\pi^2$.
The solid line is drawn at ${g}_A=1.26$ and 
the dashed line is drawn at ${g}_A=0$.
}
\end{figure}

\section{Results}

We first take a stock of
the axial charge $g_A^{0+}$ of the 
ground-state positive-parity nucleon,
which is well known and can be the references.
The axial charge $g_A^{0+}$
of the positive parity nucleon is shown in Fig.~\ref{AxialVectorC}
as open squares.
The axial charge of the positive parity nucleon
shows little quark-mass dependence,
and they lie around the experimental value 1.26.

We finally show the axial charges of the negative-parity nucleon
resonances in Fig.~\ref{AxialVectorC}.
One finds at a glance that 
the axial charge $g_A^{0-}$
of $N^*(1535)$ takes quite small value,
as $g_A^{0-}\sim {\mathcal O}(0.1)$
and that even the sign is quark-mass dependent.
While the wavy behavior might come from
the sensitiveness of $g_A^{0-}$ to quark masses,
this behavior may indicate that
$g_A^{0-}$ is rather consistent with zero.
These small values are not the consequence
of the cancellation between u- and d-quark contributions.
The u- and d-quark contributions to $g_A^{0-}$
are in fact individually small~\cite{Takahashi:2007ti}.
On the other hand,
the axial charge $g_A^{1-}$ of $N^*(1650)$
is found to be about 0.55,
which has almost no quark-mass dependence.
The striking feature is that
these axial charges, $g_A^{0-}\sim 0$ and $g_A^{1-}\sim 0.55$,
are consistent with naive nonrelativistic quark model
calculations~\cite{Nacher:1999vg,Glozman:2008vg},
$g_A^{0-}= -\frac19$ and $g_A^{1-}= \frac59$.
Such values are obtained 
if we assume that the wave functions of $N^*(1535)$ and $N^*(1650)$ are 
$|l=1, S=\frac12\rangle$ and $|l=1, S=\frac32\rangle$
neglecting the possible state mixing.
(Here, $l$ denotes the orbital angular momentum
and $S$ the total spin.)

In the chiral doublet model~\cite{DeTar:1988kn,Glozman:2007jt},
the small $g_A^{N^*N^*}$ is realized
when the system is decoupled from the chiral condensate
$\langle \bar \psi \psi \rangle$.
The small $g_A^{0-}$ of $N^*(1535)$ then
does not contradict with the possible and attempting scenario,
the 
{\it chiral restoration scenario in excited hadrons}~\cite{Glozman:2007jt}.
If this scenario is the case, the origin of mass 
of $N^*(1535)$ (or excited nucleons) is essentially
different from that of the positive-parity
ground-state nucleon $N(940)$,
which mainly arises from the spontaneous chiral symmetry breaking.
However, the non-vanishing axial charge of $N^*(1650)$
unfortunately gives rise to doubts about the scenario.

\section{Conclusions}

We have performed 
the first lattice QCD study of the axial charges $g_A^{N^*N^*}$
of $N^*(1535)$ and $N^*(1650)$, with two flavors of
dynamical quarks employing the renormalization-group improved gauge
action at $\beta$=1.95 and the mean-field improved clover quark action
with the hopping parameters, $\kappa$=0.1375, 0.1390 and 0.1400. 
We have found the small axial charge $g_A^{0-}$ of $N^*(1535)$,
whose absolute value seems less than 0.2 
and which is almost independent of quark mass,
whereas the axial charge $g_A^{1-}$ of $N^*(1650)$ is found to be about 0.55.
These values are consistent with 
naive nonrelativistic quark model predictions,
and could not be the favorable evidences for 
the chiral restoration scenario in (low-lying) excited hadrons.
Further investigations on the axial charges
of $N^*(1535)$ or other excited baryons will cast light on the
chiral structure of the low-energy hadron dynamics and 
on where hadronic masses come from.

\section{acknowledgments}
All the numerical calculations were performed
on NEC SX-8R at RCNP and CMC, Osaka University, 
on SX-8 at YITP, Kyoto University,
and on BlueGene at KEK.
The unquenched gauge configurations
employed in our analysis
were all generated by CP-PACS collaboration~\cite{AliKhan:2001tx}.
We thank L.~Glozman, D.~Jido, S.~Sasaki, and H.~Suganuma
for useful comments and discussions.
This work was supported by a Grant-in-Aid for
Scientific research by Monbu-Kagakusho
(No. 20028006 and 20540265), the 21st Century COE
``Center for Diversity and University in Physics'',
Kyoto University and 
 Yukawa International Program for 
Quark-Hadron Sciences (YIPQS).

\end{document}